\documentclass[twocolumn]{aa}
\usepackage{graphicx}
\usepackage{txfonts}
\begin{document}

\newcommand {\flux} {{$\times$ 10$^{-11}$ erg cm$^{-2}$ s$^{-1}$}}
\def\magcir{\raise -2.truept\hbox{\rlap{\hbox{$\sim$}}\raise5.truept
\hbox{$>$}\ }}

\title{{Using the ROSAT Catalogues to find counterparts for the second IBIS/ISGRI Survey Sources}}
\subtitle{}

\author{ J. B. Stephen\inst{1}, L. Bassani\inst{1}, A. Malizia\inst{1}, A. Bazzano\inst{2}, P. Ubertini\inst{2}, A.J. Bird\inst{3}, A.J. Dean\inst{3}, F. Lebrun\inst{4}, R. Walter\inst{5}}

\offprints{stephen@bo.iasf.cnr.it}

\institute{IASF/CNR, Via Piero Gobetti 101, I-40129 Bologna, Italy, 
\and  IASF/CNR, Via del Fosso del Cavaliere, I-00133 Roma, Italy,
\and  University of Southampton, Southampton, UK,
\and  CEA-Saclay, DSM/DAPNIA/Service d'Astrophysique, 91191 Gif-sur-Yvette Cedex, France,
\and  INTEGRAL Science Data Centre, Chemin d'Ecogia 16, 1291 Versoix, Switzerland}

\date{Received / accepted}

\titlerunning{}
\authorrunning{J.B. Stephen et al.}

\abstract{
The second IBIS/ISGRI survey has produced a catalogue containing 209 hard X-ray sources visible down to a flux limit of around 1 milliCrab. The point source location accuracy of typically 1-3 arcminutes has allowed the counterparts for most of these sources to be found at other wavelengths. In order to help identify the remaining objects, we have used the cross-correlation recently found between the ISGRI catalogue and the ROSAT All Sky Survey Bright Source Catalogue. In this way, for ISGRI sources which have a counterpart in soft X-rays, we can use the much smaller ROSAT error box to search for identifications. For this second survey, we find 114 associations with the number expected by chance to be $\sim$2. Of these sources, 8 are in the list of unidentified objects and, using the smaller ROSAT error boxes, we can find tentative counterparts for five of them. We have performed the same analysis for the ROSAT Faint Source Catalogue, finding a further nine associations with ISGRI unidentified sources from a total of 29 correlations, and, notwithstanding the poorer location accuracy of these sources and higher chance coincidence possibility, we have managed to find a counterpart for another source. Finally, we have used the ROSAT HRI catalogue to search the ISGRI error boxes and find 5 more X-ray objects, of which two are neither in the bright or faint source catalogues, and for which we have managed to find optical/near infrared associations. This makes a total of 19 objects with X-ray counterparts for which we have found possible identifications for nine, most of which are extragalactic.

}

\maketitle


\section{Introduction}
Stephen et al. (\cite{Stephen}) (hereafter S1) have shown that there is a strong correlation between the source catalogue from the first INTEGRAL IBIS/ISGRI survey (Bird et al. \cite{Bird}) and the ROSAT All Sky Survey Bright Source Catalogue (RASSBSC, Voges et al. \cite{Voges} ): they calculate that there are 75 associations with the number expected by chance to be almost zero. They have further provided a restricted error box for 10 unidentified ISGRI sources and indications for optical associations for most of these sources. A second ISGRI catalogue has recently been prepared and accepted for publication (Bird et al. \cite{Bird2}) which represents an extension both in exposure and sky coverage with respect to the first. This new survey contains 209 soft gamma-ray sources of which 150 have been identified with known galactic or extragalactic objects; the remaining 59 either lack identification or still have an uncertain nature. Herein we extend the correlation analysis of S1 to the new catalogue and, further, perform the same correlation with the ROSAT Faint Source catalogue. Finally we search the ROSAT HRI (short) catalogue for any other soft X-ray counterparts for still unidentified ISGRI sources. 

\section{Method}
Following S1, we have calculated the number of ISGRI sources for which at least one RASSBSC object is within a specified distance out to a maximum of around 3 degrees. For control, we then created a list of `anti-ISGRI' sources mirrored in Galactic longitude and latitude and the same correlation algorithm applied. Figure 1 shows the results of this process. As in S1, it is clear that the `anti-source'-RASSBSC distribution can be totally explained by chance while a correlation is evident in the ISGRI population. 

\begin{figure}
\centering
\includegraphics[width=8.0cm,height=10.0cm]{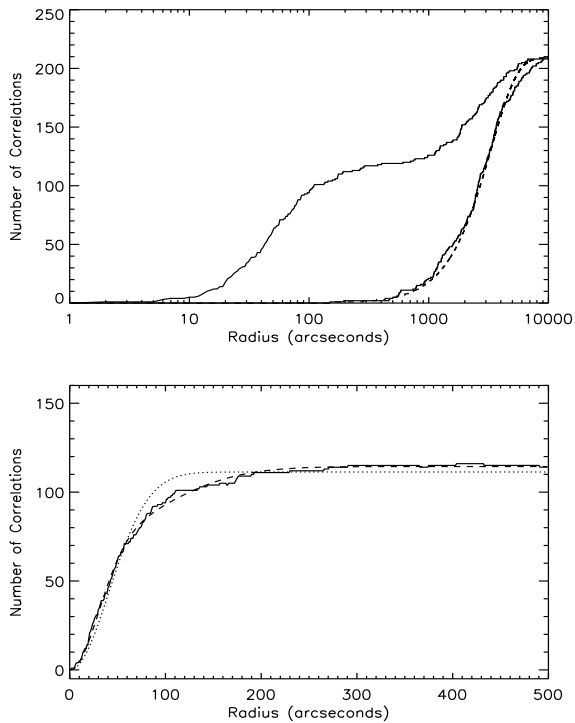}
\caption{({\em Top}) The Number of ISGRI/RASSBSC (upper solid) and `Anti-ISGRI'/RASSBSC (lower solid) associations as a function of distance. The dashed line shows the number of correlations expected by chance.({\em Bottom}) A close-up of the ISGRI-ROSAT Correlation distribution. The one-Gaussian (dotted) and two-Gaussian (dashed) fits to the data are also shown.}
\end{figure}

The detailed shape of the correlation function, as pointed out in S1, indicates not only the total number of associations (where the curve flattens) but also the point source location accuracy of ISGRI (as that of the ROSAT Bright Sources is much smaller in comparison). In contrast to the results in S1 where the distribution was reasonably well-fit with either a single or a double Gaussian function, here the single gaussian PSF is no longer compatible and a 2-Gaussian fit is required by the data. This is probably due to the improved cleaning algorithms used in producing the second survey source list which allows weaker sources to be identified which naturally have a poorer location accuracy. This fit results in a total of 114 associations, and an ISGRI PSLA of 40 arcseconds (90\%) for sources stronger than about 40 $\sigma$ and 1.8 arcminutes for weaker sources. Of these 114 sources, we would expect to find two by chance alone.

\begin{figure}
\centering
\includegraphics[width=8.0cm,height=7.0cm]{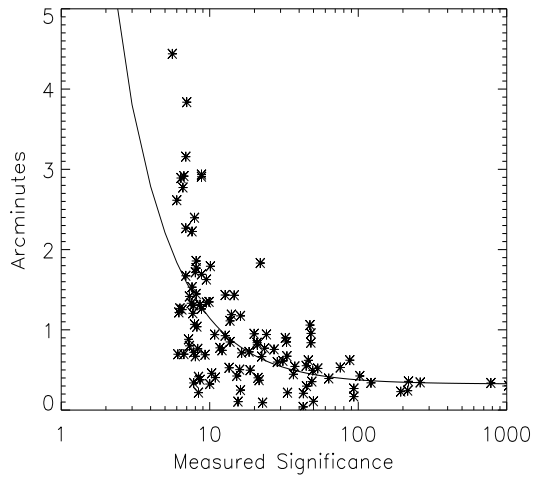}
\caption{The distribution of ISGRI-RASSBSC distances as a function of ISGRI significance.}
\end{figure}

Another manner of investigating the PSLA is to plot the ISGRI-ROSAT separation as a function of source significance. This can be seen in figure 2, which is fit (taking into account the non-gaussian nature of the distribution of distances at any one significance) by the function $\delta x = 13 S^{-1.2} + 0.33$. This can be compared to figure 6 in Bird et al.(\cite{Bird2}) where all {\em identified} sources in the catalogue have been used to create a similar distribution using the SIMBAD positions thereby obtaining the relationship $\delta x = 11.6 S^{-1.36} + 0.51$. The  width of the correlation function leads us to consider possible associations out to a distance of 4\arcmin.5, where the possibility of an RASSBSC object being within this distance of an ISGRI source is only 1.7\%.
\begin{figure}
\centering
\includegraphics[width=8.0cm,height=10.0cm]{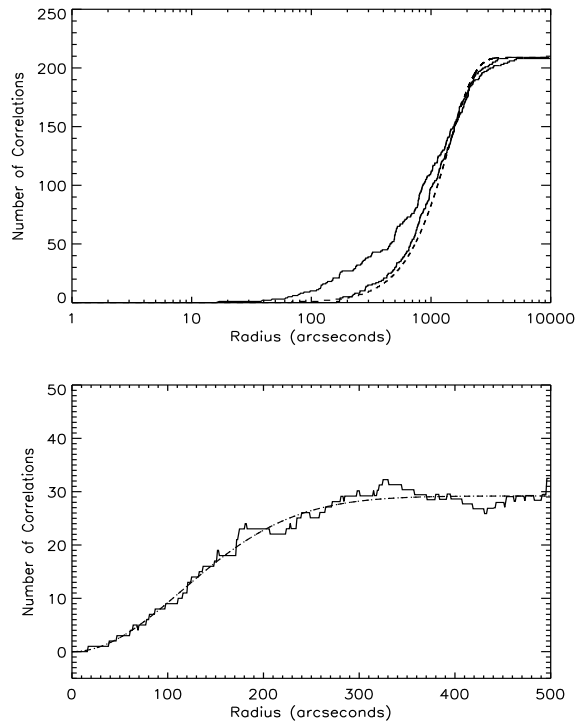}
\caption{The ISGRI-ROSAT correlation as shown in Figure 1, but using the ROSAT Faint Source Catalogue. In this case a single Gaussian fit is sufficient and a double-Gaussian is not required by the data.}
\end{figure}
From the 114 associations found, 8 correspond to sources which are in the list of unidentified or unclear identification and which have not been already discussed in S1. One of these sources, IGR J16377$-$6423, has been optically identified in the second IBIS catalogue but it is discussed here as its angular distance from the optical counterpart is greater than the ISGRI positional uncertainty. In order to increase the number of associations we have also searched the ROSAT HRI (short) catalogue for objects inside the ISGRI error boxes. We use the short catalogue as this contains a small number of bright sources thereby limiting the possible chance associations which would occur by using the full HRI catalogue. We find 5 associations of which 2 are new, resulting in a total of 10 unidentified ISGRI sources for which there are bright ROSAT counterparts. For all these sources, we report in Table 1 the relevant ISGRI and RASSBSC parameters. For the HRI positions, the typical uncertainty is 4 - 10 arcseconds. Conservatively we use 10\arcsec as the error radius in our search for counterparts as discussed in section 3.
\begin{table*}
\begin{center}
\caption{Unidentified ISGRI Sources with a RASSBSC counterpart}
\begin{tabular}{ccccccc}
\hline
&&&&&&\\ 
{\bf ISGRI} & {\bf RASSBSC}     & {\bf RASSBSC}      & {\bf RASSBSC-ISGRI}  & {\bf RASSBSC}        & {\bf HRI Coord.} & ID$^{\ast}$\\ 
{\bf Name}  & {\bf Coordinates} & {\bf Error (\arcsec)} & {\bf Distance (\arcmin)} & {\bf strength (c/s)} &                  &    \\
\hline
\hline
IGR J07597$-$3842& 07 59 41.0  $-$38 45 36.5  &   9       & 1.9 &  0.30$\pm$0.03  & 07 59 43.5  $-$38 42 36.0 & Y\\
IGR J16194$-$2810& 16 19 33.6  $-$28 07 36.5  &   8       & 2.6 &  0.19$\pm$0.02  & 16 19 33.0  $-$28 07 38.1 & T\\
IGR J16377$-$6423& 16 38 18.3  $-$64 21 07.0  &   9$^{a}$ & 4.4 &  1.24$\pm$0.07  &                           & Y\\
IGR J16482$-$3036& 16 48 15.5  $-$30 35 11.0  &  13       & 1.3 &  0.15$\pm$0.02  &                           & Y\\
IGR J16500$-$3307& 16 49 55.1  $-$33 07 13.0  &  14       & 1.3 &  0.08$\pm$0.02  &                           & N\\
XTE1716$-$389    & 17 15 57.7  $-$38 51 52.0  &  12       & 2.9 &  0.64$\pm$0.05  & 17 15 56.6  $-$38 51 50.5 & T\\
IGR J17513$-$2011& 17 51 13.0  $-$20 12 14.5  &  13       & 1.3 &  0.07$\pm$0.02  &                           & N\\
IGR J17544$-$2619& 17 54 28.3  $-$26 20 35.0  &  23       & 0.7 &  0.08$\pm$0.02  &                           & Y$^{b}$\\
IGR J18483$-$0311&                            &           &     &                 & 18 48 17.3  $-$03 10 18.0 & T\\
2E1853.7$+$1534  &                            &           &     &                 & 18 56 00.4  $+$15 37 57.5 & Y\\
\hline
\end{tabular}
\end{center}

$^{a}$ source extended in ROSAT image\\
$^{b}$ source identified by Gonzales-Riestra et al. \cite{Gonzales}\\
$^{\ast}$ Y/N = Yes/No, T=Tentative
\end{table*}
We have also performed the same correlation analysis with the ROSAT Faint Source Catalogue (RASSFSC), containing a total of 105924 sources (Voges et al. \cite{Voges2}). The results are shown in Figure 3. It is clear that a correlation exists, but is much weaker. A single Gaussian is a good fit to the correlation function giving a 90\% width of 2.6 arcminutes. The total number of correlations is 29, but because of the large number of RASSFSC sources we would expect anything up to 17 chance coincidences. Within 4\arcmin.5 there are 27 possible associations in comparison with 8 expected by chance. Of these 27, 9 correspond to ISGRI sources which remain unidentified and they are listed in Table 2 with their relevant parameters.

However, in the following discussion, we restrict ourselves to consider the six possible associations which have distances of $\le$2.5 arcminutes where we would only expect 2/27 to be by chance (or less than one in the table). In the case of IGR J17488$-$3253 we know that the most likely counterpart is a RASSBSC object at $\sim$1\arcmin which has been fully discussed in S1. 

\section{Results}

In the following we give details on the search for counterparts and highlight a number of candidates. We use the following catalogues: USNO B1 in optical (Monet et al. \cite{Monet}); 2MASS in the near infrared (Cutri et al. \cite{Cutri}); IRAS in the far infrared (Joint IRAS Science Working Group \cite{IRAS}) and NVSS for radio (NRAO/VLA sky Survey, Condon et al. \cite{Condon}). In a few cases, data available in the literature or in archives are sufficient to permit source identification.\\
 
{\bf IGR J07597$-$3842}
This is the only case where we find two bright ROSAT sources possibly associated with the ISGRI object. Both RASSBSC sources have an HRI detection, although the HRI positions are offset by around 1\arcmin.6 in declination in both cases. However since the HRI instrument provides a better position than the RASSBSC, we assume the coordinates given by this instrument to be more accurate. The association of these two HRI sources with the ISGRI object is fully discussed by Den Hartog et al. (\cite{Hartog}) who concluded that one of the two being located 3\arcmin.3 from the ISGRI position (RA(2000)=07 59 59.1  and Dec (2000)=$-$38 42 39.0) is a less likely candidate counterpart and therefore is not reported in Table 1. Further support to this conclusion comes from  detection of the source quoted in Table 1 in the Beppo-SAX wide field cameras data at a location compatible with the HRI one. Den Hartog and co-author also report detection of this source at radio and far-infrared frequencies. Further to this report, Molina et al. (\cite{Molina}) localized the optical/near-infrared counterpart of this X-ray/radio source and also indicated an extragalactic origin on the basis of the near-infrared photometry. In view of all these broad band  characteristics and location away from the galactic plane, we conclude  that this is  a new active galactic nucleus, which has so far eluded detection.\\
 
{\bf IGR J16194$-$2810}
This ROSAT source, also detected by the HRI instrument as 1RXH J161933.0$-$280738, has been associated with the HST Guide Star catalogue object GSC6806.00016 (B $\sim$ 13.6-14.0, Voges et al. \cite{Voges}); also known as USNO B1 0618$-$0432633, this object is also fairly bright at near-infrared wavelengths with 2MASS J-H-K magnitudes of 8.3, 7.3 and 7 respectively. Although we consider this to be the most likely association, it is important to note that while this is the only near-infrared counterpart present within the HRI error box, 3 more objects are listed in the USNO-B1 catalogue but all with lower magnitudes.\\
 
\begin{table*}
\begin{center}
\caption{Unidentified ISGRI Sources with a RASSFSC counterpart}
\begin{tabular}{cccccc}
\hline
&&&&&\\
{\bf ISGRI} & {\bf RASSFSC}     & {\bf RASSFSC}      & {\bf RASSFSC-ISGRI}  & {\bf RASSFSC}  & ID$^{\ast}$ \\ 
{\bf Name}  & {\bf Coordinates} & {\bf Error (\arcsec)} & {\bf Distance (\arcmin)} & {\bf strength (x10$^{-2}$ c/s)}&\\
\hline
\hline
IGR J07565$-$4139       & 07 56 18.8  $-$41 38 23.5  &   17 & 2.0     & 4.26$\pm$1.37 & T\\
IGR J09026$-$4812       & 09 02 38.4  $-$48 14 08.0  &   20 & 2.5     & 2.71$\pm$0.90 & N\\
IGR J11114$-$6723       & 11 11 04.8  $-$67 24 01.5  &   15 & 2.0     & 1.93$\pm$0.90 & N\\
4U1344$-$60             & 13 47 38.4  $-$60 37 00.5  &   23 & 1.4     & 5.48$\pm$1.71 & Y\\
IGR J15359$-$5750       & 15 35 52.8  $-$57 50 55.0  &   19 & 1.0     & 3.34$\pm$1.50 & N\\
IGR J17488$-$3253$^{a}$ & 17 48 35.9  $-$32 55 13.5  &   13 & 2.9     & 5.38$\pm$1.81 & N\\
IGR J18048$-$1455       & 18 04 41.2  $-$14 56 49.5  &   17 & 3.0     & 3.37$\pm$1.36 & N\\
IGR J18214$-$1318       & 18 21 29.0  $-$13 16 41.5  &   26 & 2.5     & 2.36$\pm$1.07 & N\\
IGR J18259$-$0706       & 18 25 57.5  $-$07 10 21.5  &   11 & 4.0     & 3.36$\pm$1.17 & N\\
\hline
\end{tabular}
\end{center}

$^{a}$ more likely associated with a RASSBSC (S1)\\
$^{\ast}$ Y/N = Yes/No, T=Tentative\\
\end{table*}

{\bf IGR J16377$-$6423}
The ROSAT source is likely associated with a bright X-ray cluster of galaxies from the CIZA (Clusters in the Zone of Avoidance, Ebeling et al. \cite{Ebeling}) survey: CIZA 1638.2$-$6420 also known as Triangulum Australis (see figure 4). This cluster at z=0.058 has a ROSAT luminosity of the order of 10$^{45}$ erg s$^{-1}$. It has also been reported in the recent RXTE slew survey catalogue (XSSJ16384$-$6424), which clearly indicates emission up to about 20 keV (Revnivtsev et al. \cite{Revnivtsev}). Zimmerman et al. (\cite{Zimmerman}) associated the X-ray source to the central galaxy of the cluster, ESO101$-$G004, on the basis of the X-ray contours that suggested that most of the observed X-ray emission was coming from the galaxy itself and not from the extended group or cluster gas component.  The IBIS source is slightly shifted (by 4\arcmin.4) from the ROSAT position but still compatible with the cluster extension in X-rays (see figure 4). It is possible that the high energy object is related to a galaxy cluster member or to a hot spot in the cluster itself.\\

\begin{figure}
\centering
\includegraphics[width=8.0cm,height=8.0cm]{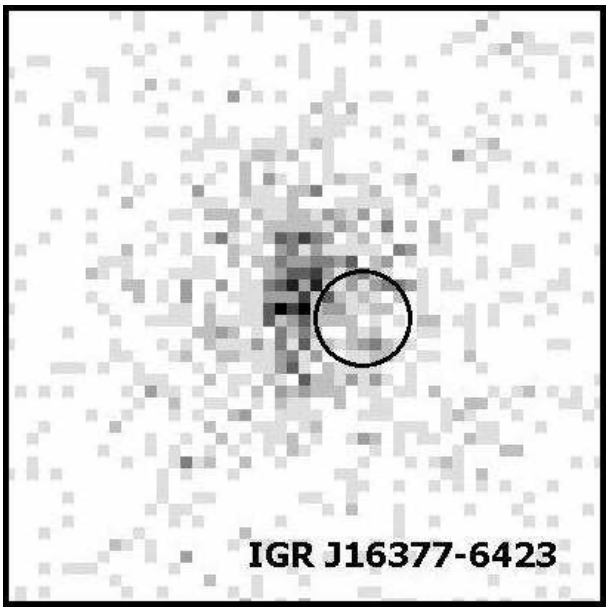}
\caption{RASSBSC Image of the region containing IGR J16377$-$6423 showing the ISGRI 3\arcmin radius error box}
\end{figure}

{\bf IGR J16482$-$3036}
Compatible with  the ROSAT error box, we find an NVSS source at  RA(2000)=16 48 16.6 and  Dec(200)=$-$30 35 07.5 (about 15\arcsec away) with a 20cm flux of 3.5 mJy. In NED this object coincides with a 2MASS extended object, associated with a galaxy. The optical/infrared counterpart is fairly bright with  B=15.82, R= 13.38, J=13.94  and K= 12.18. In view of the infrared morphology and the radio and X-ray detection, we suggest that this is a background active galaxy viewed through the galactic plane.\\
 
{\bf IGR J16500$-$3307}
Within the ROSAT error box of IGR J16500$-$3307 there are too many (8) USNO B1 counterparts for a fruitful identification search, a clear indication that an  error radius greater than 10 arcsec makes the identification procedure very difficult. However in the  near-infrared only two counterparts are found in the 2MASS catalogue of which only one is also listed in the DENIS survey (http://cdsweb.u-strasbg.fr/denis.html). This object (DENIS J164955.6$-$330701) is located at RA(2000)=16 49 55.64 and Dec(2000)=$-$33 07 01.9 and has B=16.9, R=16, J=14.3 and K=13.75. The other infrared source (RA(2000)=16 49 55.17 and Dec(2000)=$-$33 07 08.9) is much weaker (J=16.26 and K=15.55) and is not detected at optical wavelengths. Clearly we need a smaller error box (e.g. by Chandra or XMM observations) in order to pinpoint the true counterpart.\\  
 
{\bf XTE1716$-$389}
This ROSAT source, also detected by the HRI instrument as 1RXH J171556.7$-$385150, has been associated with the still unidentified X-ray source XTE J1716$-$389 (also KS1716$-$389). Although considered as a transient (Remillard \cite{Remillard}), this object has been detected by the TTM/COMIS telescope onboard Mir-Kvant (Emilyanov et al. \cite{Emilyanov}), by ASCA (see HEASARC archive) and is also reported in the EXOSAT Slew survey catalogue as EXO J171557.7$-$385 (Reynolds et al. \cite{Reynolds}). All these observations suggest that XTE J1716$-$389 is more likely to be a persistent object although highly variable. This is confirmed by the analysis of the XTE/ASM light curve and dedicated XTE/PCA observational data, which further indicate the possible presence of a period of $\sim$ 100 days as well as dips of 40 days duration in the light curve (Wen et al. \cite{Wen}). This indicates that the object is possibly a binary system. Within the HRI positional uncertainty, we find only one optical candidate at RA(2000)=17 15 56.8 and Dec(2000)=$-$38 51 56.3 (B=18.9, R=17.8) and a few near-infrared objects; the only optical source found is also in the 2MASS catalogue with J=14.9, H=13.3 and K=12.8.\\
 
{\bf IGR J17513$-$2011}
At least 6 objects are present in the USNO-B1 catalogue within the ROSAT positional uncertainty: the R magnitudes of these objects are all greater than 17.0 ; five of them are also detected in the 2MASS catalogue. Inspection of the NVSS sky image archive indicates emission at the 2 mJy level, although no radio source is reported within the ROSAT error circle; it is possible that this emission is due to a source at the NVSS detection limit or just noise in the radio map.\\
 
{\bf IGR J17544$-$2619}
The identification and nature of this source is fully discussed by Gonzales-Riestra et al. (\cite{Gonzales}) who report on an XMM observation of the ISGRI error box. This observation localizes the soft X-ray counterpart of the ISGRI source and further locates its infrared and optical counterpart.  It is an extremely interesting case from the point of view of ISGRI/RASSBSC association as the XMM data excludes the association of the ISGRI source with the ROSAT object, i.e. the ROSAT object is at the border of the XMM error box. Either this is one of the few spurious associations expected or the ROSAT/XMM sources are the same object with the ROSAT positional uncertainty underestimated.\\

{\bf IGR J18483$-$0311}
This is one of the two objects in our sample only detected by the HRI instrument. Within the X-ray positional uncertainty two sources are found in the USNO-B1 catalogue: the first located at RA(2000)=18 48 16.9 and Dec(2000)=$-$03 10 21.7 has R=19 and I=17.7 while the second with coordinates RA(2000)=18 48 17.2 and Dec(2000)=$-$03 10 16.5 (R=19 and I=15.3) is also detected at near-infrared frequencies with J=10.74, H=9.296 and K=8.46.\\

{\bf 2E1853.7$+$1534}
This source also has only an HRI detection and two optical counterparts within the X-ray error box: the first located at RA(2000)=18 56 00.6 and Dec(2000)=$+$15 37 58.4 (B= 18.7 and R=15.7) is also detected at near-infrared frequencies with J=13.6, H=12.6  and K=11.4 while the second with coordinates RA(2000)=18 56 00.8, Dec(2000)=$+$15 38 00.4 is dimmer (B=19.3 and R=7.1) and is not seen in the 2MASS catalogue. Compatible with the HRI position we also find an NVSS source (3.4 mJy at 20 cm) with a positional uncertainty of 6.5\arcsec which allows the exclusion of the second source as a likely counterpart. In NED this object coincides with a 2MASS extended source, again associated with a galaxy. The infrared morphology, the radio and X-ray detection and the location just above the galactic plane all suggest that this is another active galaxy.\\

{\bf 4U1344$-$60}
Within the ROSAT error box of this bright historical source we find an object recently detected by XMM (1XMM J134736.1$-$603704); the much smaller positional uncertainty of XMM allows us to pin point a near-infrared  counterpart at  RA(2000)=13 47 36.0 and Dec(2000)=$-$60 37 03.8 (J= 14.0, H=12.3, K= 11.1) but not any optical source, most likely due to absorption in the galactic plane. This source is also detected at far-infrared frequencies but is not detected in radio due to its location outside the sky region covered by the NVSS. It has been suggested that this is a low-galactic low-redshift active galaxy (Michel et al. \cite{Michel}).\\

For the rest of the sources in Table 2, within their error boxes there are too many optical and/or near-infrared counterparts in the USNO-B1/2MASS catalogues for a fruitful identification search. However, we notice that close to the IGR J07565$-$4139 ROSAT position (at 42\arcsec and within the RASSFSC 2$\sigma$ error box) there is a 2MASS extended object (2MASX J07561963$-$4137420) classified as a galaxy and also detected at far-infrared wavelengths; if this is a true association then this is another extragalactic object as concluded above for IGR 16482$-$3036 and 2E1853.7$+$1534.

\section{Conclusions}
The use of various ROSAT catalogues has allowed us to restrict the size of positional error boxes associated with nineteen unidentified INTEGRAL IBIS/ISGRI sources allowing at least tentative identifications for nine of them, of which most turn out to be extragalactic in nature. For the remaining objects even the ROSAT error box is not sufficiently small so as to allow unambiguous identification and follow-up observations by Chandra and/or XMM are solicited.

\begin{acknowledgements}
We acknowledge the following funding: in Italy the Italian Space Agency via contracts I/R/046/04;in UK via PPARC grant GR/2002/00446; in France CNES for support during ISGRI development and INTEGRAL data analysis. This analysis has made use of the HEASARC archive which is a service of the Laboratory for High Energy Astrophysics (LHEA) at NASA/ GSFC
and the High Energy Astrophysics Division of the Smithsonian Astrophysical Observatory (SAO), of the SIMBAD database, operated at CDS, Strasbourg, France and of NASA/IPAC Extragalactic Database (NED) which is operated by the Jet Propulsion Laboratory, California Institute of Technology, under contract with the National Aeronautics and Space Administration. 
\end{acknowledgements}

\end{document}